# A comparative analysis of the geometrical surface texture of a real and virtual model of a tooth flank of a cylindrical gear

Jacek Michalski[*], Leszek Skoczylas[**]
Rzeszów University of Technology, Faculty of Mechanical Engineering and Aeronautics,
W. Pola 2, 35-959 Rzeszów, Poland

**Abstract**

The paper presents the methodology of modelling tooth flanks of cylindrical gears in the CAD environment. The modelling consists in a computer simulation of gear generation. A model of tooth flanks is an envelope curve of a family of envelopes that originate from the rolling motion of a solid tool model in relation to a solid model of the cylindrical gear. The surface stereometry and topography of the tooth flanks, hobbed and chiselled by Fellows method, are compared to their numerical models Metrological measurements of the real gears were carried out using a coordinated measuring machine and a two – and a three-dimensional profilometer. A computer simulation of the gear generation was performed in the Mechanical Desktop environment.

[*]Corresponding author. Tel.: (48-17) 8651570.
E-mail address: jmichals@prz.rzeszow.pl (J. Michalski)
[**]Corresponding author. Tel.: (48-17) 8651258.
E-mail address: lsktmiop@prz.rzeszow.pl (L. Skoczylas)

## 1. Introduction

The variety and multitude of the application of gears results in their remarkable role in machines and devices. Although there is a tendency to use electric machines and electronic systems in drivers and machine tool control in particular, gears still find widespread application. The surface stereometry and topography of gear tooth flanks and their research have always been a major technological problem and are decisive for the operational value of gears.

The paper compares the surface stereometry and topography of modelled tooth flanks of cylindrical gears obtained by a three dimensional simulation of gear generation with the geometrical surface structure (GPS) of real gears obtained through hobbing and chiselling by Fellows method.

## 2. Literature review

While making use of the envelope curve condition and the operational continuity of the surface contact of the tool and the generated tooth, a system of equations in their open parametric form is aimed at. The equations are useful for tool setting [1,2,3,4,5,6]. In the generation method, the profile of the tooth flank [7,8,9] and its surface [3,4,5,10,11] are determined analytically and numerically as an envelope of a curve family formed by the profile of the tool blades in rolling motion with the generated gear. Matrix, vector and differential equations are used for a tooth flank description. They also make it possible to describe complicated geometric shapes of gear flanks. Papers [12, 13] give a mathematical model of the tooth flank generation of a pinion and a cone gear of circular-arched tooth line where the size and localization of the cooperation mark as well as the character of the plot and the deviation level of movement irregularity have been predetermined.

A model of the stereometric shape of the gear tooth flank and its surface topography can be defined also in the CAD environment. The method is based on the procedure of logical taking off the solids representing the tool and the generated object. In the CAD environment it is possible not only to obtain a complete model of gear teeth but also simulate the generation process [11,12,13,14,15]. Simulations of cylindrical [11,14,15], cone [12,13,14] and worm gears were carried out. The cutting edge of a tool that is computer modelled can have arbitrarily defined profile. The literature [16] gives mathematical equations (1) and (2) which enable calculating tooth profile deviation $\delta_y$ and helix deviation $\delta_x$ (Fig. 1). They result from the nature of hobbing.

(a) (b)

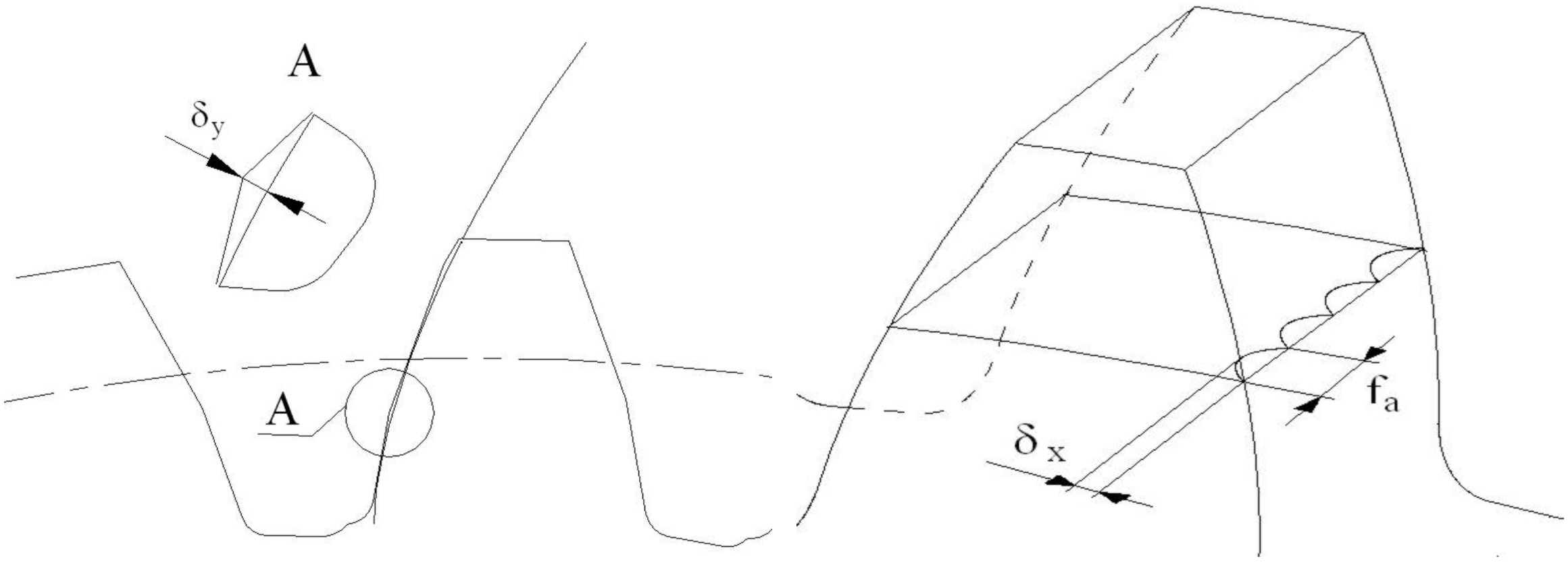

Fig.1. Tooth profile precision deviation $\delta_y$ (a), helix deviation $\delta_x$ and valley spacing on helix $f_a$ (b).

The denotations used in them are as follows:

$\alpha_n$ teeth profile angle, $d_0$ tool pitch diameter, $f_a$ feed per revolution, $m_n$ normal gear module, $z_1$ number of hob coils, $z_2$ number of gear teeth and $n_i$ number of hob blades.

$$\delta_x = \tan\alpha_n \cdot \left( \frac{d_0}{2} - \sqrt{\frac{d_0^2 - f_a^2}{4}} \right) \qquad (1)$$

$$\delta_y = \frac{\pi^2 \cdot z_1^2 \cdot m_n \cdot \sin\alpha_n}{4\, z_2 \cdot n_i^2} \qquad (2)$$

Equations (1) and (2) allow calculating one value characterizing describing the whole tooth flank. Actually there are value differences for the deviations between the head and the root of the tooth. [15]. However, in paper [17] the calculated experimental roughness profile of tooth flank surfaces measured along the tooth profiles does not show distinct temporary roughness changes that might be a representation of the hob blades taking successive positions at particular moments of the generation.

Making modern gears necessitates carrying out a number of trials, analyses and measurements [18, 19, 22]. The research includes the semi-finished product phase [22], the stereometry perfection [17,19,20,24,25,26] and its final formation taking into account the effectiveness and cost of production [18]. Despite an increase in the production cost, the number of hard material gears [23,24,26, 27, 28] keeps growing. Their hobbing is usually done with a cutting edge of relatively big diameter, often of multicoil structure. In the diagonal method, the cutting edge is extra long. In the way of reduction and finishing generation, gear shaving in a hard material, honing and grinding are also applied for 2%, 8% and 12% of gear rims, respectively.

The autocorrelation function is used to theoretically describe GPS and analyse it qualitatively, whereas the power spectral density function is applied to the quantitative analysis [29, 30]. The autocorrelation function is more useful for the estimation of random surfaces, whereas the spectral density function is more appropriate in the case of determined surfaces [31, 32, 33]. Consequently, a geometrical surface structure analysis of gear tooth flanks requires the use of the two functions. The height parameters of surface topography are very helpful for predicting the use of the surfaces, which are totally separated from each other with an oil film, for hydrodynamic lubrication [32, 33, 35, 36]. In the case of dry and mixed lubrication, due to lack of straight relations between GPS features and tribological functions [34, 35, 36]. The GPS correctness of the gear tooth flanks is evaluated by means of height, curvature, density of peaks or summits [31, 34], slope [36, 37, 38, 39, 40, 41] or plasticity index [31, 34]. The calculation of the latter is often based on the slope of the surface profile [31, 33, 34]. Since slope is not an internal surface characteristic, it should be considered in various scales [38]. A three dimensional surface topography can be described with a set of 14 parameters, with an extension to 17 [32, 37]. Anizotrophies of the surface are described with the arithmetic mean of the absolute values of the slope $P\Delta a$, $R\Delta a$ and the quadratic mean of the slope $P\Delta q$, $R\Delta q$, of the characteristic orientation of the machined surfaces. The slope value is calculated from two, three or seven profile points [33]. The surface anizotrophy of machine elements is also evaluated with a relative change of the maximum and minimum variance of the ordinate values of the surface orientation profiles $K_\alpha$ [42]. The value of the parameter square Pq (Rq) represents an approximated predicted value of the variance of ordinate values. The combination of the slope growth of the profile or surface segments, the so called conregular model, enables the estimation of the anizotrophy or asymmetry of the surface [43]. The variation of the profile slope shows not only the amplitude distribution but also demonstrates frequency behaviour. The distribution of the material volume in the roughness height area is characterized by the parameters of its description by means of the following methods: the secand method [31, 32, 33], probability method [44] and that of the least slope and ordinate of the point of the least slope of the standardized material curve approximated with a function of three parameters [45]. The stereometric measurements of the surface are carried out with a coordinate measuring machine [31]. The measurement of the topography of the tooth flanks is carried out with great accuracy and in a short time with profilometers and microscopes. GPS flaws and defects can also be analysed [29, 30, 31, 46, 47].

**3. Characteristic of the gear subjected to tests**

The semi-finished toothed wheels were made by die forging. The material was alloy, low-carbon steel for carbonizing of the AMS 6265 sort. The blank was of 30-35 HRC. The envelopes of the wheels were thermally toughened up to 35÷41 HRC. The cylindrical wheel with straight teeth had two 6.3 mm and 5.1 mm rims and a 1.814 mm module. The toothed rim, of an 88.9 mm pitch diameter, was chiselled by Fellows method. The other toothed rim of a 156.0297 mm pitch diameter was out-cut hobbed without parallel feed of the cutter axis. The generation was carried out with a Pfauter PE 300 hobbing machine made by American Pfauter Limited (Estes, Illinois, USA). A monolithic, single right-curling cutter of ASP 2040 steel, 70 mm diameter and 14 Lorenz made flutes were applied. The tool flank and tool face of the cutter edges was covered with a layer of titanium nitride. Ordinary, moderate peripheral speed of the hob was 140 m/min, axial feed was 2 mm/rev and two passes with two 3.7 mm and 0.1 mm depth cuts. The coolant was 30 compounded oil (absolute viscosity at 40°C, 120 $mm^2/s$). Radial run-out of the cylindrical control surfaces of the hob and the cylindrical surface of the generated wheel rim did not exceed 0.01 mm. A 0.2 mm tooth grinding allowance was left for the tooth side to be ground. The hob ensured undercutting at the roof (protuberances) of the profile of the tooth flanks.

Gear generation chiselling was carried out in three passes by Fellows method, using an LS186CNC slotter (Lorenz Ettlingen, Buchholz-Mendt, Germany). The pot-type cutter, also a Lorenz product, with 56 teeth, was made of ASP 2030 steel and covered with a layer of titanium nitride (TiN). The in-feed method, with revolution feed, of 3.4, 0.5 and 0.1 mm incision depth was applied. The in-feed was 0.003 mm/double cutter pitch and revolution feed for three passes was respectively 0.285, 0.228 and 0.104 mm/double cutter pitch. Consequently, the number of the double cutter pitches on the scale between adjacent teeth was 20, 25 and 55. The speed of the to-and-pro motion of the cutter was 20, 25 and 30 m/min, respectively. The cutting fluid was Ferrocol EB oil (absolute viscosity at 40°C, 7 $mm^2/s$). The radial runout of the control cylindrical surface of the cutter and the cylindrical surface of the generated gear rim did not exceed 0.01 mm.

It follows from the generation parameters that making the whole profile engages the cutter edges 28 times whereas a pinion cutter edges necessitate 116 contacts. It results from the engagement factor.

## 4. Methodics of tooth flank modelling

Modelling tooth flanks in the CAD environment makes use of commands of turning, copying, shifting and taking away the drawn solids of the wheel and the tools. The reciprocal turning of the toothed wheel and the tool is not smooth but rather stroke-like in character. The interdependence between a wheel revolution and the shift of the tool is described with equation (3). The designations are as follows: b tangent shift of the tool corresponding to the wheel revolution by angle φ (°); $d_t$ turning diameter of the toothed wheel.

$$b = \frac{\pi d_t \varphi}{360} \qquad (3)$$

The simulation procedure in the CAD environment, written as a macrodefinition, is as follows:

**rotate**

gearwheel {revolution of generated gear}

x, y {coordinates of the centre of wheel revolution}

φ {value of wheel revolution angle (Fig. 2)}

**move**

tool {tool shift corresponding to a wheel revolution}

0, 0 base point of the shift

b, 0 {value of the shift (Fig. 2)}

**copy**

| | |
|---|---|
| tool | {tool copy and its placement on the turning diameter of the wheel $d_t$} |
| 0, 0 | {base point} |
| 0, a | {location of the tool copy (Fig. 2)} |
| **subtract** | |
| gear wheel | |
| tool | {subraction of the dipped tool solid volume from the gear wheel model} |
| **rscript** | {restarting the procedure}. |

The kinematics of the hobbing and chiselling simulation was presented in Fig. 2.

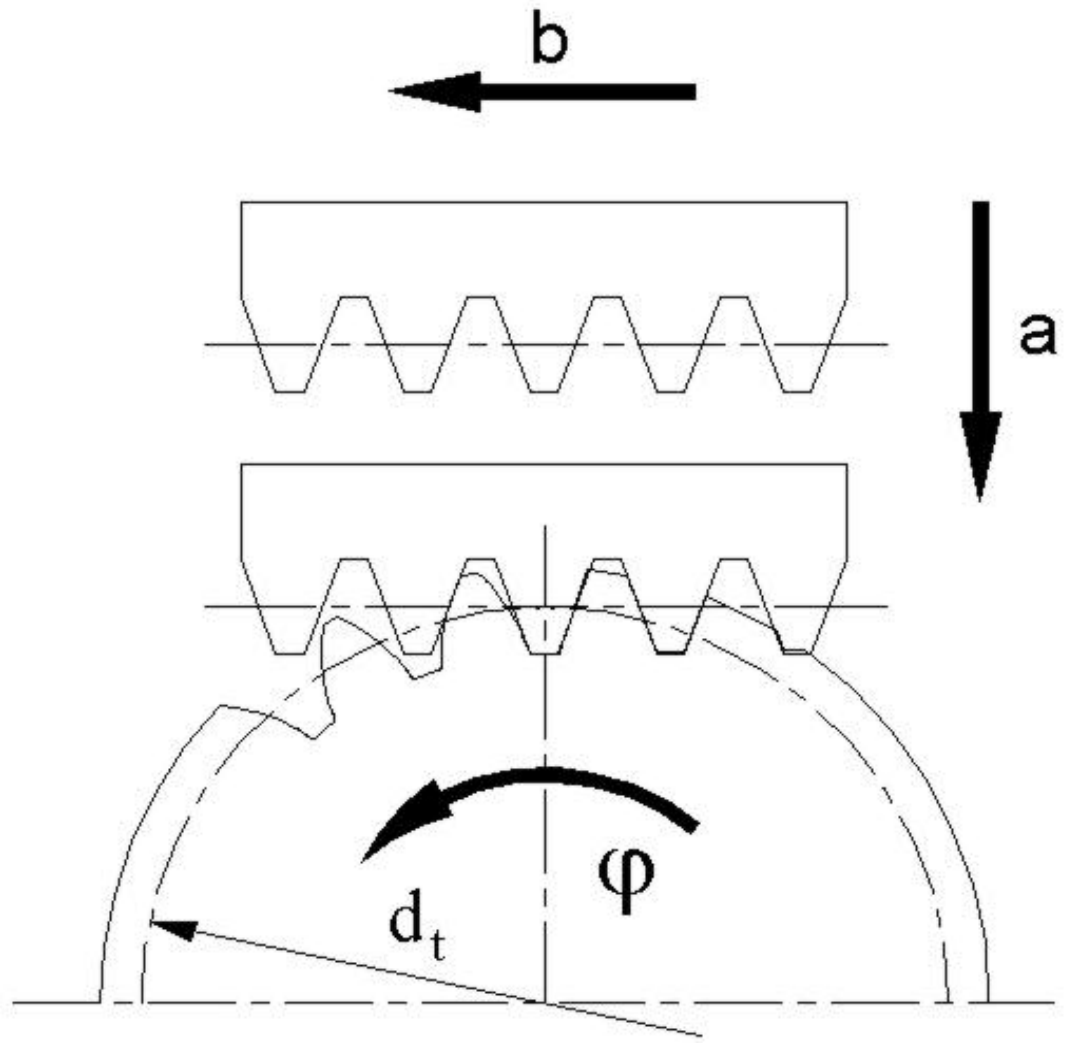

Fig.2. Straight tooth cylindrical gear generation simulation includes: push of tooth reference cylinder of hob by “a” which ensures tangency of tooth pitch plane of reference cylinder to pitch cylinder of diameter $d_t$, hob feed b with a simultaneous wheel turn by angle φ from equation (3).

In the case of hobbing (Fig. 3a), the tool is a model of a plain milling cutter. The computer simulation of the gear generation takes place in a three dimensional environment. A tooth profile is formed due to the turning of the gear wheel and the tool. The helix results from the hob blade moving along the wheel axis. The feed motion of the tool (b) is dependent on the rotary motion of the generated gear wheel (φ) in accordance with the transmission ratio of the technological gear, equation (3). Shift c corresponds to the value of the axial feed per wheel turn. A computer simulation of gear generation by chiselling was carried out in a similar way (Fig. 3b).

(a) (b)

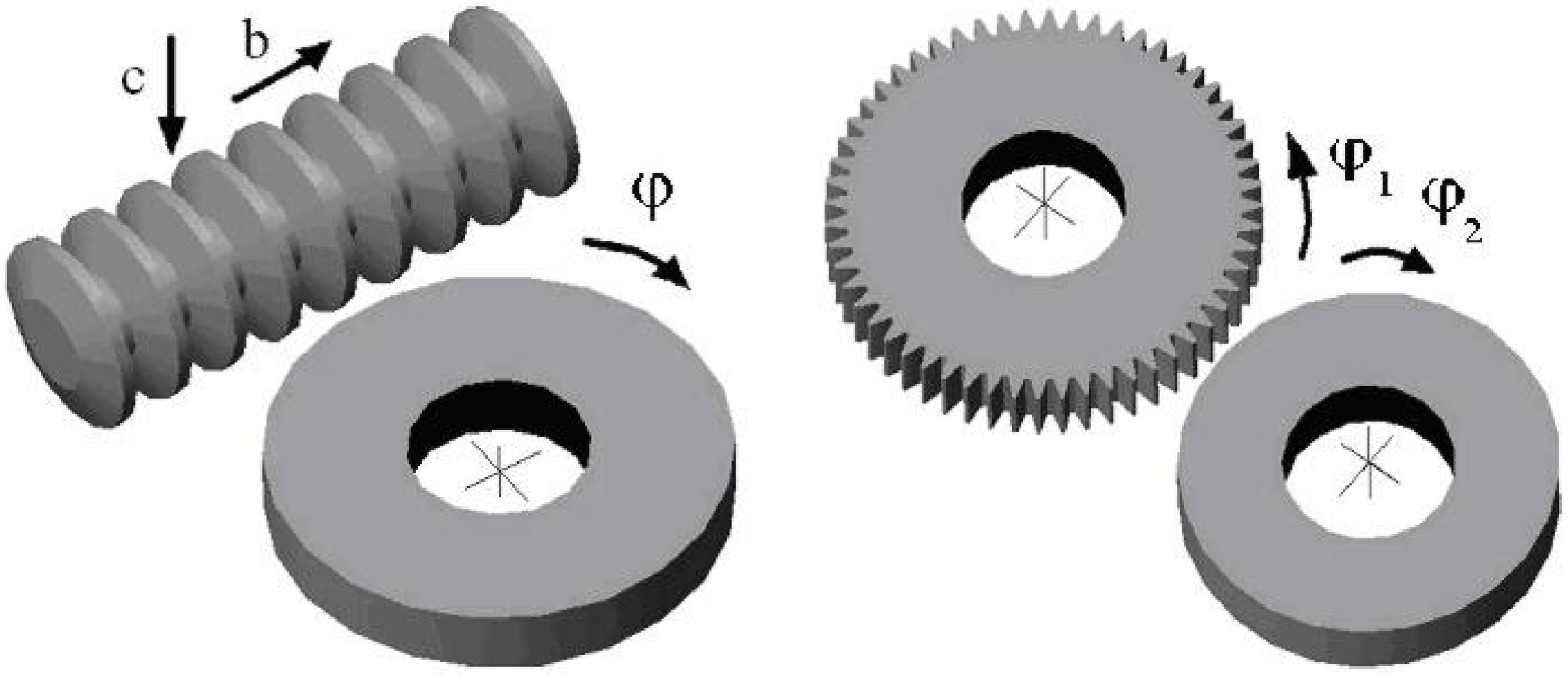

Fig.3. Simulation of straight tooth cylindrical gear generation with hobbing (1) and Fellows chiselling (b) including pushes b, c and turns $\varphi$, $\varphi_1$ and $\varphi_2$.

The tool in this case is a solid model of a pinion cutter. The turn of the tool by angle $\varphi_1$ is related to the wheel turn $\varphi_2$, in accordance with the transmission ratio of the technological gear. During the construction of the model, a simplification of the solid model of the hob and its angular position in the course of the simulation was adopted. The hob model has no helix so it does not have to be turned by the spiral angle. Examples of the flanks of the teeth obtained through hobbing and chiselling simulation is presented in Fig. 4.

(a) (b)

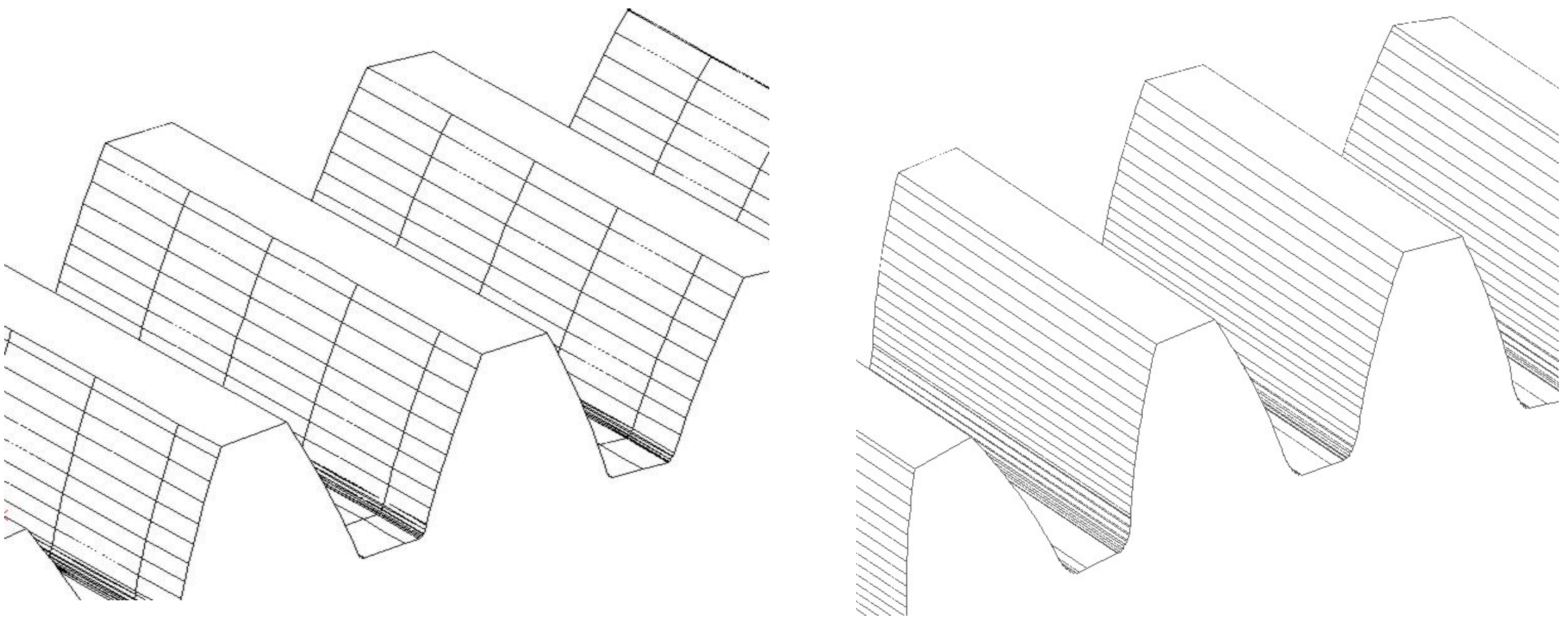

Fig.4. Tooth flanks of a toothed wheel obtained with hobbing (a) and Fellows chiselling simulation (b).

In the first case the flanks of the toothed wheel are composed of a number or regularly distributed surfaces corresponding to the successive layers of the material being removed by the tool blades. The surfaces are concave, therefore the tooth flanks are scaly in character. It is evidenced by a curvature analysis of the tooth flanks. Their depth along the height of the teeth varies. The flanks of the teeth, chiselled by Fellows method, are also composed of a number of concave surfaces. They have concavities only along the profile along the tooth height (Fig. 4b). They run rectilinearly, tangentially to the helix.

Taking the simulation parameters corresponding to the real generation parameters, it is possible to obtain a solid gear wheel model without the errors resulting from the real generation process [19, 21, 27, 28]. To obtain the models of the surfaces of the gear teeth of the constructional parameters, the following methodics of calculating the parameters of a computer simulation of gear generation was adopted:

a) for a hobbed wheel, the tool a hobbing cutter has 14 flutes on the perimeter. The angle of the circular pitch is 4.186 (°). Due to this, one stroke of the cutting edge of the tool corresponds to an angular displacement equal 0.299 (°).

b) For a chiselled gear wheel, the tool a pinion modular cutter makes 55 double strokes on the circular pitch. The angle of the pitch of the toothed wheel is 7.347 (°). Because of this, one stroke of the tool corresponds to 0.134 (°) of the wheel turn. Assuming that the tool has 56 teeth, its angular displacement per one stroke is 0.117 (°). Applying the presented commands of the generation programme simulation, three - dimensional models of the tooth flank surfaces of cylindrical gear wheels were obtained.

## 5. The course of the research

The stereometric shape of the tooth flanks of the gear wheels was measured with a CNC coordinate measuring machine, PNC model, Klingelnberg Sohne (Remscheid, Germany). The stylus ended with a spherical surface of a 1mm radius. The GPS measuring base was an imaginary axis formed by the surfaces of the centre holes on the faces of the gear wheel shaft. The measured coordinates and the software offered by the producer of the measuring machine made it possible to compute single pitch deviation $f_{tp}$, total cumulative pitch deviation $F_p$, total profile deviation $F_a$, total helix deviation $F_\beta$, tooth thickness variation $R_s$ and runout $F_r$ (Table 1). The surface topography of the tooth surface was examined with a Stemi 2000 C. Zeiss stereoscopic microscope through a viewing system consisting of coupled cameras (CCD). The results were processed with the Matrox Intellicam programme. Then, a measurement of the topography of the tooth surface of the gear wheels in the middle of their width was carried out. A Talyscan 150 Taylor Precision (Leicester, GB) three dimensional profilometer was used. For the tooth space flanks getting into and out of machining the square measuring area was 3 mm. 360000 results were obtained (600 lines, 600 measuring points each). The sampling step and the spacing of the measuring tracks were 5 μm. Measurements on the head and root of the tooth in a square of a 1.50 mm side were also made. The lengths of the steps and measuring track spacing were the same as before. The scanning speed was 2 mm/s in each case. The calculation of the surface topography parameters was carried out with the Maintains Map Universal programme, Digital Surf programme (Tables 2 and 3, Figs 5 and 6). In addition, basing on the results obtained, a topography analysis was carried out using our own, original software. Extra measurements of the tooth flank surfaces with a Form Talysurf Series 2 Taylor Pneumo profilometer were made. Along the tooth height, the measuring length was 3 mm, along the helix it was 4 mm. The probing step was 0.25 μm. The gauging points of the applied profilometers were conical, of a 90° apex angle, and ended with a 2.5 μm radius spherical surface. The profile parameters were found using both the producer's and our own original software. The results are shown in Table 4 along the tooth height and along the helix, the associated profile was formed using a 5-grade-multinominal and a straight line, respectively. The analysed length of the profile and the helix of the hobbed gear wheel and the chiselled one was 2550 μm (85 points with a 30 μm step). In the examination of all the original profiles, the short wave irregularities of the surface were not separated with a λs profile filter.

Model flank profiles of hobbed and chiselled teeth were analysed at a normal cross - section and separated with an interpolating circle (associated profile) of a 15.0844 mm radius (chiselled gear wheel) and a 26.5826 mm radius (hobbed gear wheel). The circle was determined with two points of the tooth flanks situated on the wheel rollers with the radii of 43.1868 mm and 45.6875 mm for the chiselled rim and 76.282 mm and 79.1975 for the hobbed one. The circles were found approximately, as the Mechanical Desktop pack does not include circle and cylinder approximation procedures [48]. The flanks of the model obtained through hobbing simulation had a 2763.7 μm long tooth profile (81 points with a 34.12 μm step), and a 4000.6 μm helix length (145 points with a 27.59 μm step). The flanks of the model obtained through chiselling simulation were analysed along a 2691.2 μm teeth profile height (80 points with a 33.64 μm step). Example results are shown in Figs 7 and 8. The shortest wave lengths, described with the point of the greatest curvature of the spectrum density function of cumulated power, are 70 μm and 23 μm after hobbing and chiselling, respectively (Tables 2 and 3). Their counterparts for model gear wheels have the values of 90 μm and 374 μm. The adopted probing step lengths of the topography of the gear wheel surfaces as well as their spacing were considered correct. They are smaller than half wavelength from the spectrum density function of cumulated power [32, 37].

The stereometry height of the tooth flank surfaces was described with the arithmetic mean SPa and quadratic mean SPq deviation of the surface height, total height of the surface irregularities SPt and quadratic mean height of the surface roughness summits SPqsum. The summits were calculated from 8 values of the adjacent ordinates of the square surface. The following values were found: mean material volume SPmr, mean void volume SPmvr, valley fluid retention index SPvi and core fluid retention index SPci. Roughness spacing was described with the following values: correlation length Spal wavelength of a typical gear wheel surface along width SP(1/fx) and along tooth height SP(1/fy), the determined point of the greatest curvature of the cumulated areal power spectral density. The paper also includes the calculated arithmetic mean and quadratic mean surface slope SPΔa, SPΔq. It was denoted as SPΔax for the helix and for the teeth profile it was marked along height SPΔay. Other shape parameters were also considered: the developed interfacial area ratio SPdr, density of surface summits SPds, arithmetic mean of the curvature summits SPsc (denoted as SPscy for teeth profile and as SPscx for helix); and the fractal dimension SPfd. The distribution of the surface coordinates was described with the surface skewness ratio SPsk and surface kurtosis ratio SPku. The description of the surface material ratio curve based on the secant contains: reduced summit height SPpk, reduced valley depth SPvk, core depth SPk as well as upper material ratio Sr1 and lower material ratio Sr2. The description of the relative material ratio of the surface in the probability density system includes: the values of the standard deviation of the roughness summit heights SPpq, values of the standard deviation of the valleys SPvq and the relative material ratio at the plateau to valley intersection SPmq. The approximated, standardized curve of the material ratio was described with the slope at the inflexion point dp1 and its ordinate ypp. The texture was estimated with the texture aspect ratio of the surface SPtr and the isotropy index SPizo. In view of the great variation of the surface texture direction SPtd, the paper does not give its value. The contour map and the angular plot of the areal power spectral density function were analysed.

Ordinates Z(x) of the surface roughness profiles of the examined areas of the tooth profiles with the assigned X values were subjected to the following analysis. The spectrum of the power spectral density, the plot of the standardized autocorrelation function and that of the cumulated power spectral density were found. Numerically, the variance of the roughness heights was defined with equation (4) where $Pq^2$ is the variance of the profile coordinates Z(x) within the measuring length. Maximum $P^2q$ max and minimum $P^2q$ min value refer to the profile along the tooth height or the line along the tooth width [42].

$$K\alpha = (P^2q\,\max - P^2q\,\min) / P^2q\,\max \qquad (4)$$

The vertical parameters of the surface profile were: arithmetic means Pa and quadratic means Pq of the profile ordinate deviation, total height of the profile Pt, ten point height on profile PzJIS determined from three or five elementary lengths of 0.8 mm; maximum peak height of the profile Pp, incompletness ratio of the profile Pp/Pt, skewness ratio of profile Psk and kurtosis ratio of profile Pku. Horizontal parameters were found: mean flute widths of the profile PSm elements, mean spacing of the local peaks of profile PS, wave length determined with the point of the greatest curvature of the cumulated plot of the power spectral density P(1/f), quadratic mean of profile Pλq wavelength and the correlation length from the autocorrelation function Pβ0.1 of 0.1 value. The cumulated power spectral density $G^2(\lambda)$ was also calculated. The hybrid parameters were: the arithmetic means RΔa and quadratic means PΔq, the slope of the profile and the curvature of the peaks. Three profile ordinates were used do calculate: summit curvature Pρc3, peak density Pds3 and the quadratic mean of peak heights Pδ*. Parameters Pρc3, Pds3 and Pδ*, calculated for the probing step of 1μm, are given in Table 4. The curve of the material ratio was described with: kernel profile depth Pk, reduced peak height PΔk, reduced valley depth Pvk and material ratio Mr1 and Mr2. Mean deviation of the peak surfaces Ppq, mean deviation of valley surfaces Pvq and relative material ratio Pmq were given for the material curve of the probability density of the profile.

Surface topographies, tooth profile and helix of real toothed wheels and models were also analysed after separating their geometric shape with Gauss's filters of the appropriate length. They were 2D and 1D filters. The application of the filter was due to the noticed mistakes of separating the shape of the tooth flanks with an approximating cylinder and interpolating circle. An GPS analysis covered 12 toothed wheels.

## 6. Result

The selected accuracy deviations of making cylindrical gear wheels i.e. deviations of monominal tooth flanks and runout after hobbing and chiselling are given in Table 1.

Table 1
Selected surface stereometry precision parameters of gear tooth flanks (standard deviations in brackets)

| Deviations relevant to corresponding flanks of gear teeth and run-out information | Hobbing | | Fellows chiselling | |
|---|---|---|---|---|
| | Gear drive side | Gear drive no side | Gear drive side | Gear drive no side |
| 1. Total profile deviation $F_a$ (μm) | 19.9 (2.1) | 17.5 (1.2) | 8.2 (0.9) | 7.1 (0.5) |
| 2. Total helix deviation $F_\beta$ (μm) | 8.2 (0.9) | 4.6 (0.3) | 2.4 (0.3) | 2.0 (0.2) |
| 3. Single pitch deviations $f_{pt}$ (μm) | 24.3 (2.6) | 11.3 (0.8) | 5.9 (0.6) | 4.5 (0.3) |
| 4. Total cumulative pitch deviations $F_p$ (μm) | 30.9 (3.3) | 18.0 (1.3) | 41.7 (4.4) | 39.1 (2.8) |
| 5. Pitch line run-out $F_r$ (μm) | 56.8 (6.1) | | 26.6 (2.8) | |
| 6. Maximum variation of the chordal thickness $R_s$ (μm) | 43.2 (4.6) | | 18.8 (2.0) | |

A presentation of stereometrics characteristics of the tooth point and the root area after hobbing is shown in Fig. 5. An identical characterization for toothed wheels chiselled by Fellows method is given in Fig. 6. Figures 5 and 6 include an angular plot of the function of power spectral density, accumulated power spectral density and arbitrary power spectral density along the helix and tooth profile. The lists of the roughness parameters of the analysed surfaces can be found in Tables 2 and 3. The plots of cumulated power spectral density, autocorrelation function, power spectral density spectrum and the curve of material ratio (Figs 7 and 8) were found for the tooth and helix profiles singled out from the surface topography. The results refer to the tooth flanks of the wheel model. The slopes of the roughness segments and the slope increase are demonstrated in Fig. 9 (hobbing) and Fig. 10 (chiselling). The slopes of the roughness segments were calculated from the seven ordinate values of the roughness height. The expected values of the primitive profile of the surface of the teeth getting into and out of machining, measured with a Form Talysur profilometers, are shown in Table 4. Table 5 and 6 include the parameters of the unfiltered profile and those of the surface roughness of the irregularities of tooth and helix profiles of the machined wheels and elaborated wheel models.

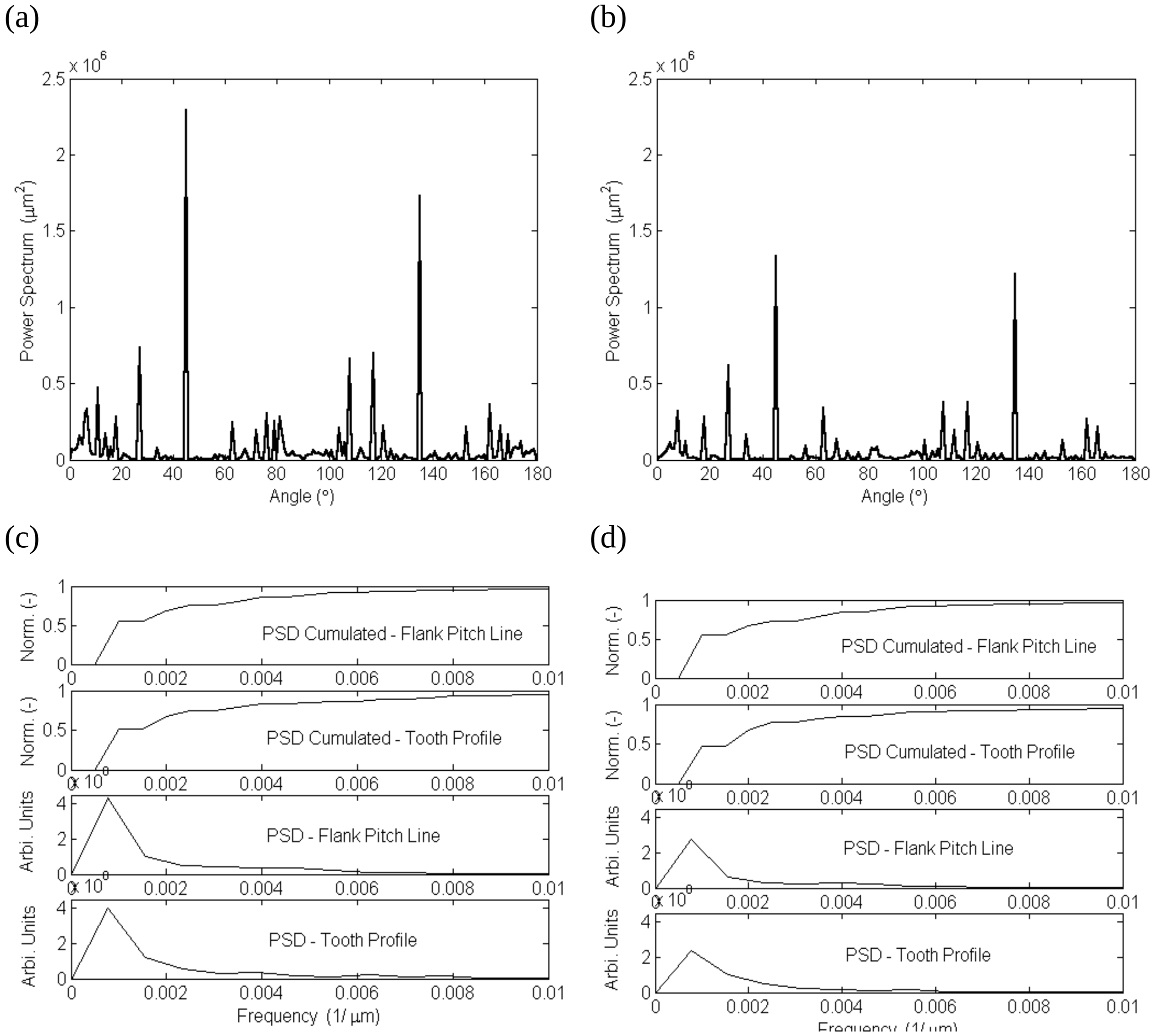

Fig.5. Characterization of root (a), (c) and tooth point surface (b), (d) of a hobbed cylindrical gear wheel for the tooth space flanks getting out of machining: (a), (b) angular plot of power spectral density, (c), (d) cumulated and arbitrary power spectral density along helix and tooth profile.

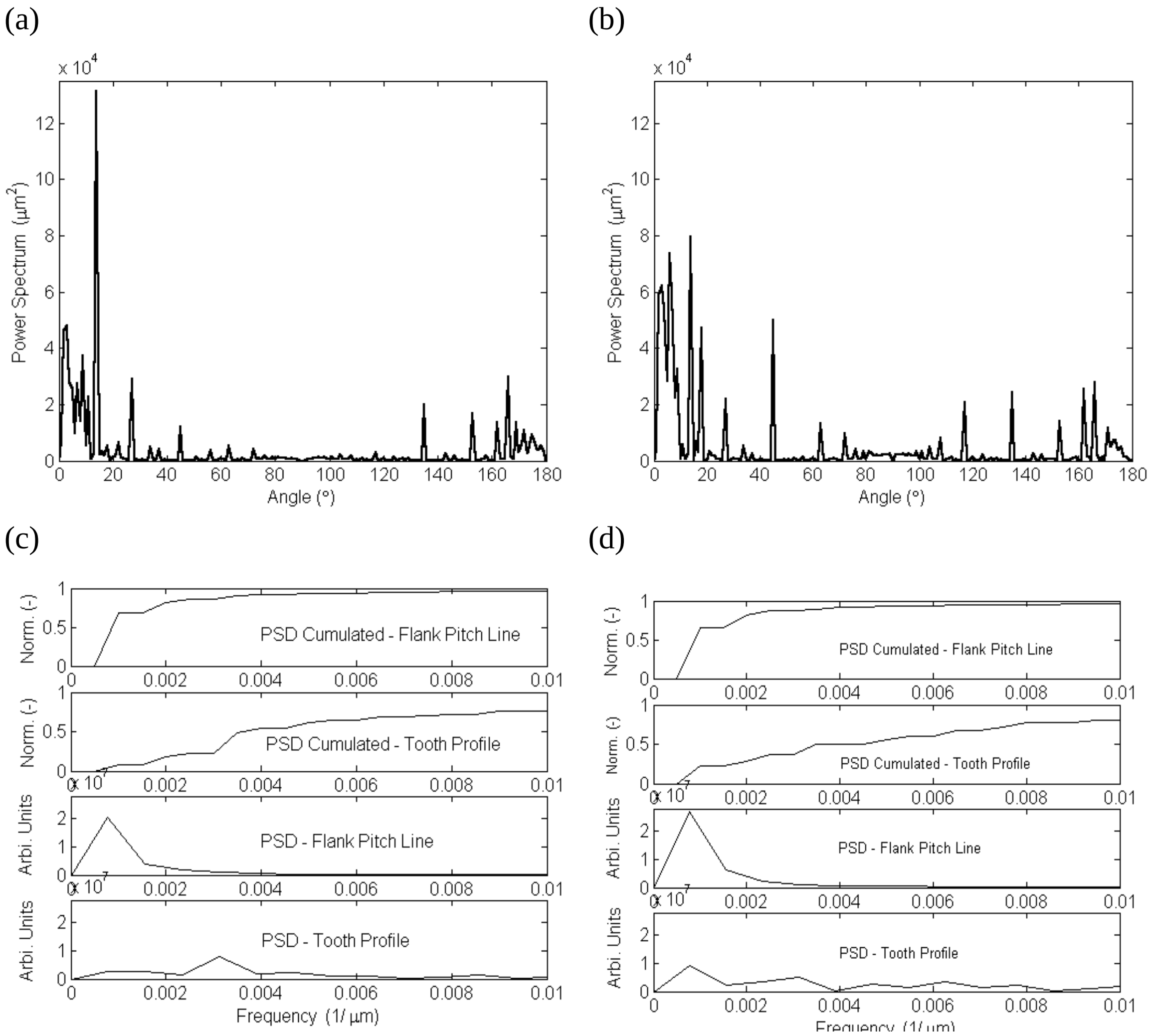

Fig.6. Characterization of root (a), (c) and tooth point surface (b), (d) of a Fellows chiselled cylindrical gear wheel for the tooth space flanks getting out of machining: (a), (b) angular plot of power spectral density, (c), (d) cumulated and arbitrary power spectral density along helix and tooth profile.

(a) (b)

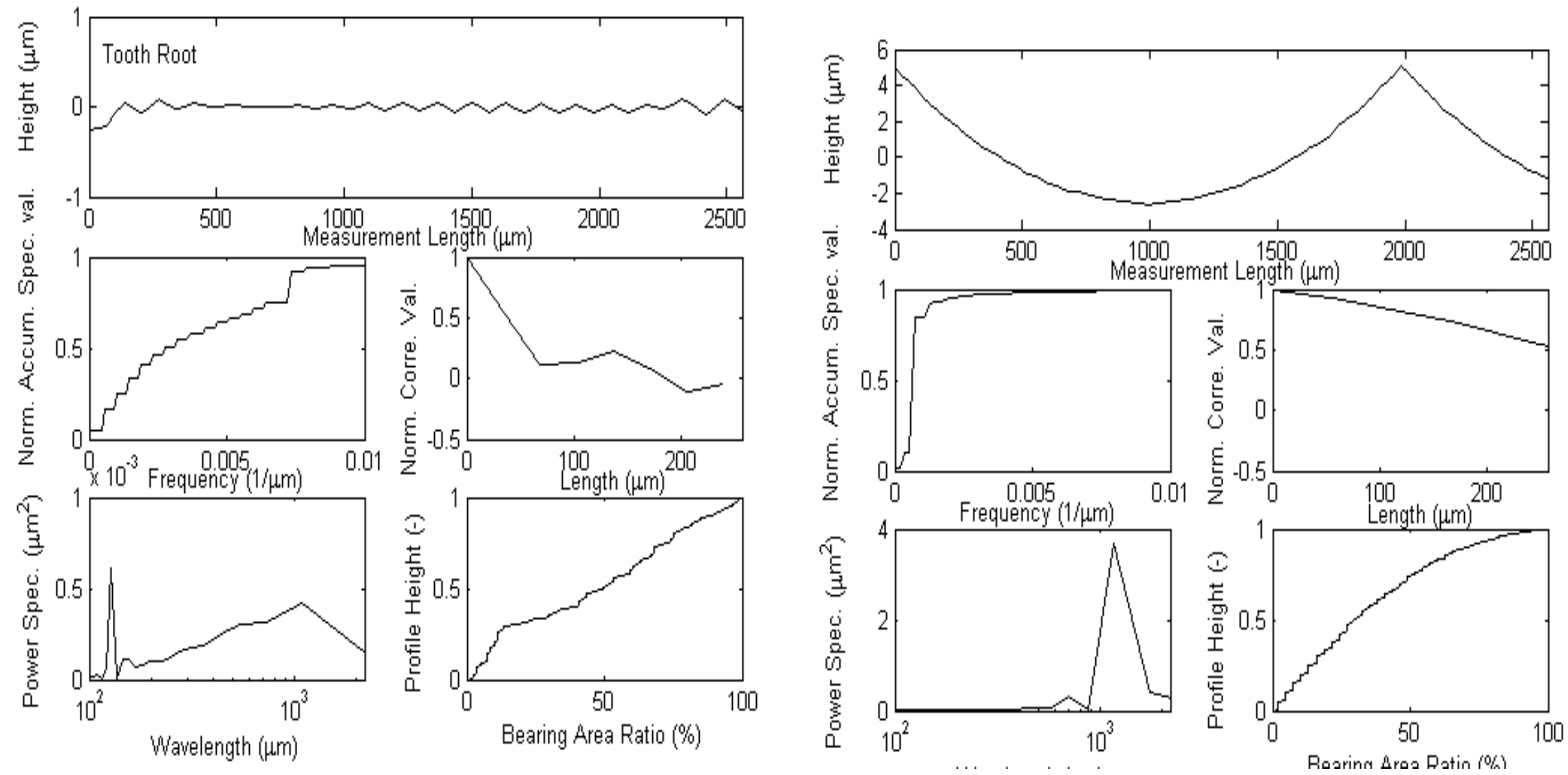

Fig.7. Roughness profile of model surface of a hobbed gear tooth flank for tooth profile (a) and helix (b) from Mechanical Desktop data. Characteristics of surface roughness profile: cumulated power spectral density, autocorrelation function, Spectrum of power spectral density, autocorrelation function, spectrum of power spectral density, and material ratio curve: Gauss's filters λc=0.8mm (a) and λc=8mm (b) were used.

(a) (b)

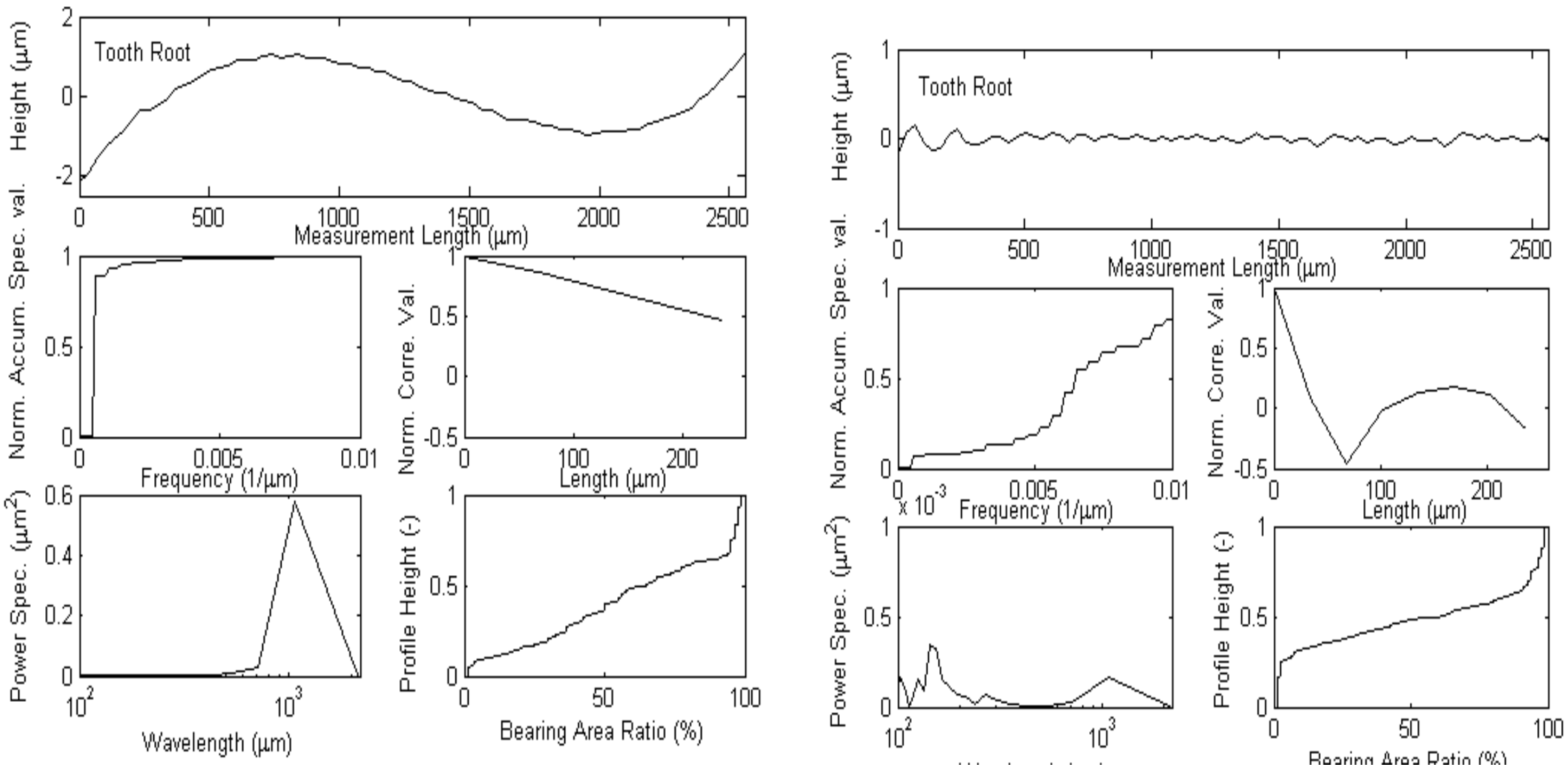

Fig.8. Primitive profile (a) and roughness profile (b) of model surface of Fellows chiselled gear tooth flank for tooth profile from Mechanical Dektop data. Characteristics of surface roughness profile: cumulated power spectral density, autocorrelation function, spectrum of power spectral density, and material ratio curve: A λc=0.8 mm Gauss's filter was used.

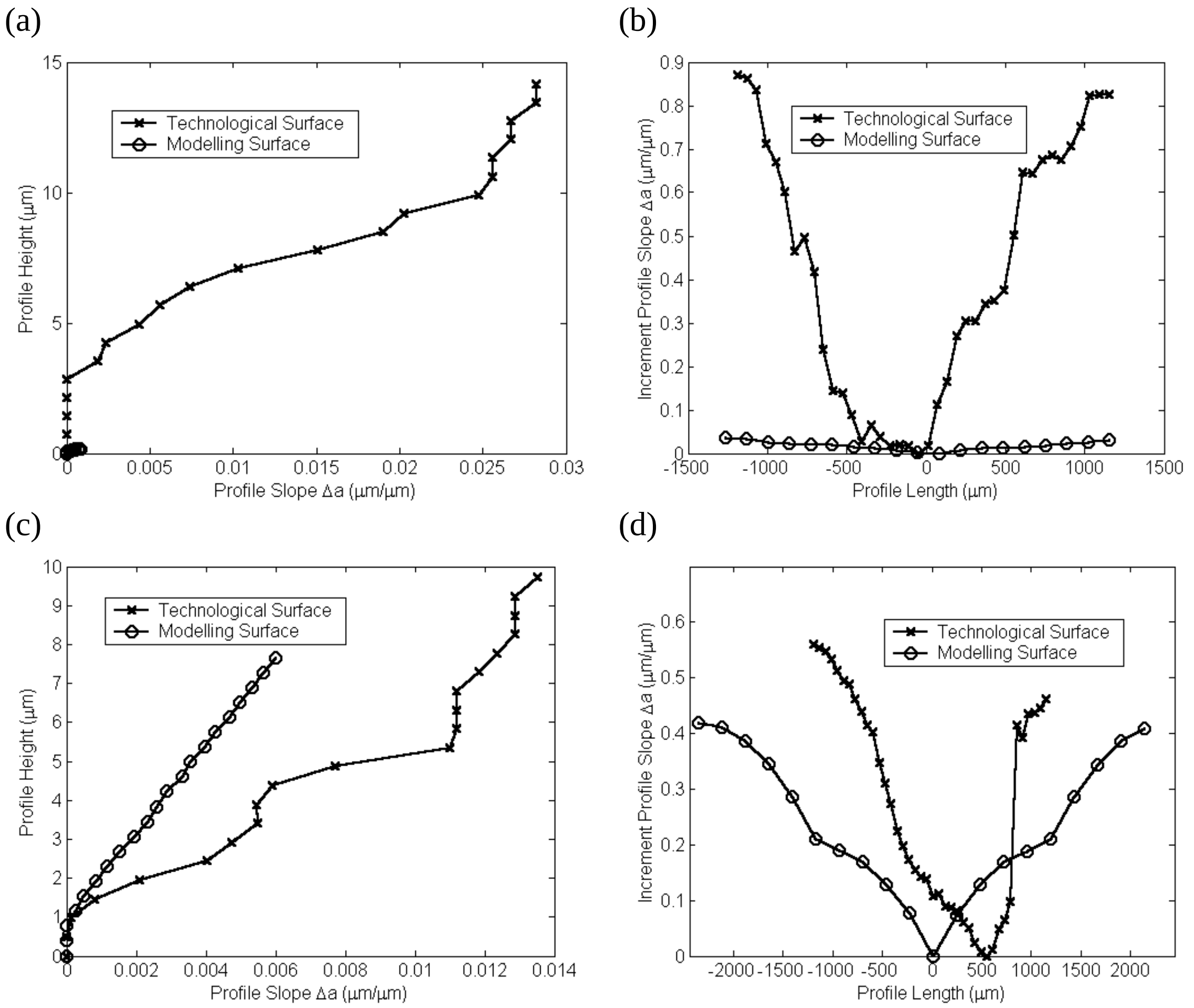

Fig.9. Characteristics of surface roughness profile slope of side out hobbed gear tooth flank along tooth height (a), (b) and along helix (c), (d). Characterization of profile height with regard to slope (a), (c). Slope increase along profile length (b), (d).

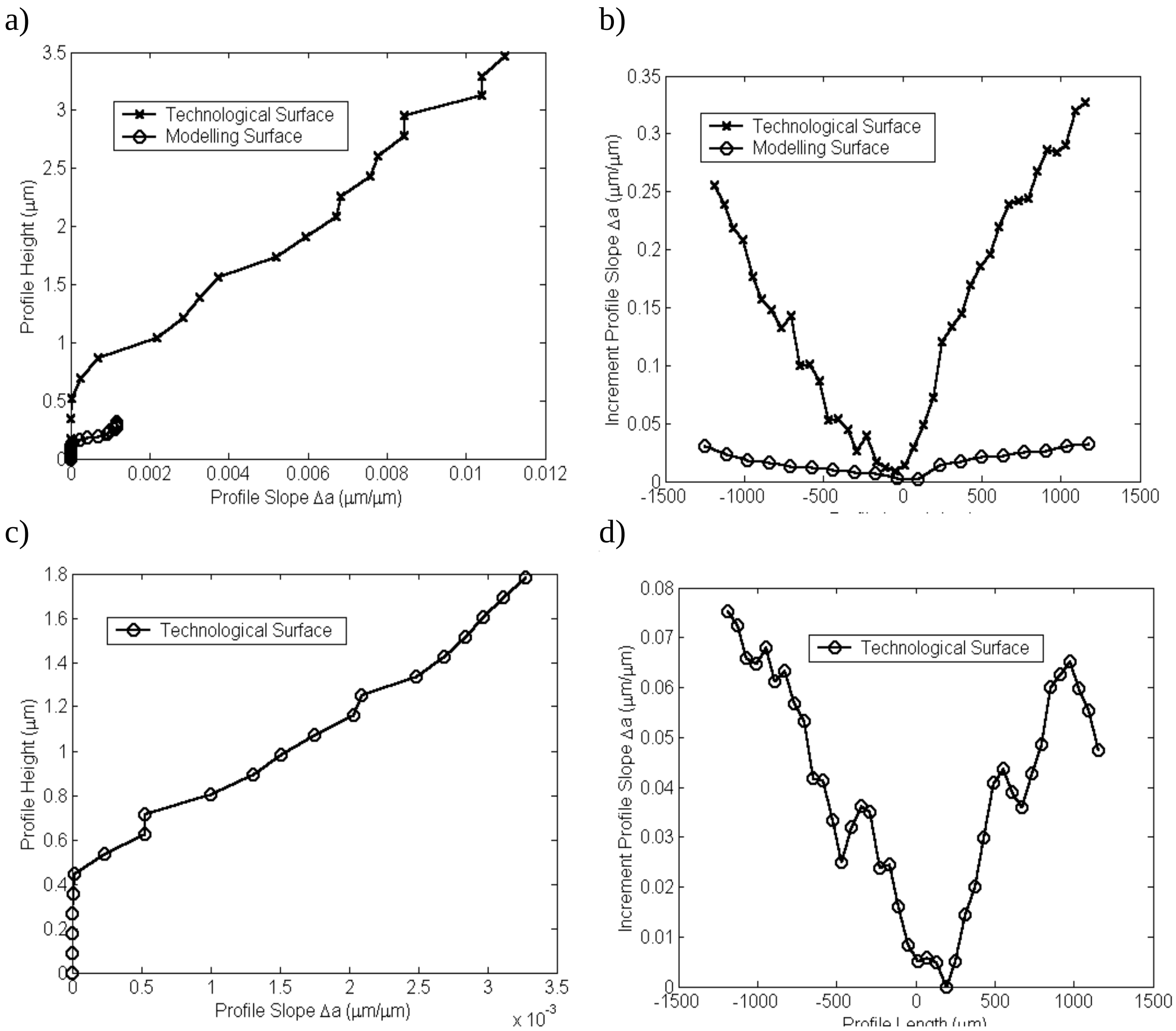

Fig.10. Characteristics of surface roughness profile slope of side out Fellows chiselled gear tooth flank along tooth height (a), (b) and along helix (c), (d). Characteristic of profile height with regard to slope (a), (c). Slope increase along profile length (b), (d).

Table 2
Surface topography parameters of hobbed gear tooth flanks

| Parameters | Gear drive side | Gear drive no side | | |
|---|---|---|---|---|
| | Whole side | Whole side | | Whole side |
| SPa (μm) | 6.49 | 4.13 | 1.80 | 3.08 |
| SPq (μm) | 7.69 | 4.93 | 2.24 | 3.88 |
| SPt (μm) | 41.8 | 31.90 | 19.15 | 21.55 |
| SPsk (-) | -0.14 | 0.22 | -0.29 | -0.01 |
| SPku (-) | 2.34 | 2.37 | 3.69 | 2.67 |
| SPmmr ($mm^3/mm^{2)}$ | 0.0196 | 0.0148 | 0.0090 | 0.0118 |
| SPmvr ($mm^3/mm^{2)}$ | 0.0222 | 0.0171 | 0.0102 | 0.0500 |
| SPds ($pks/mm^{2)}$ | 744 | 714 | 811 | 667 |
| SPtr (-) | 0.40 | 0.47 | 0.62 | 0.40 |
| SPal (μm) | 630 | 394 | 196 | 294 |
| SP isotropy (%) | 40 | 47 | 33 | 40 |
| SPfd (-) | 2.04 | 2.11 | 2.06 | 2.01 |
| SPΔq (μm /μm) | 0.0658 | 0.0641 | 0.0660 | 0.0600 |
| SPsc (1/μm) | 0.0110 | 0.0093 | 0.0092 | 0.0094 |
| SPdr (%) | 0.210 | 0.20 | 0.22 | 0.18 |
| SPbi (-) | 0.70 | 0.56 | 0.79 | 1.22 |
| SPci (-) | 1.36 | 1.62 | 1.38 | 1.56 |
| SPvi (-) | 0.10 | 0.08 | 0.13 | 0.14 |
| SP(1/fx) (μm) | 111 | 500 | 107 | 112 |
| SP(1/fy) (μm) | 133 | 333 | 97 | 108 |
| SPΔax (μm /μm) | 0.0218 | 0.0208 | 0.0199 | 0.0223 |
| SPΔay (μm /μm) | 0.0307 | 0.0286 | 0.0276 | 0.0304 |
| SPq sum (μm) | 0.11 | 0.11 | 0.16 | 0.27 |
| SPscx (1/μm) | 0.0046 | 0.0038 | 0.0040 | 0.0039 |
| SPscy (1/μm) | 0.0049 | 0.0043 | 0.0040 | 0.0044 |
| SPpq/SPt (-) | 0.18 | 0.12 | 0.13 | 0.13 |
| SPvq/SPt (-) | 0.13 | 0.19 | 0.21 | 0.23 |
| SPmq (-) | 0.05 | 0.83 | 0.71 | 0.73 |
| SPpk/SPt (-) | 0.31 | 0.23 | 0.23 | 0.24 |
| SPvk/SPt (-) | 0.22 | 0.23 | 0.28 | 0.24 |
| SPk/SPt (-) | 0.13 | 0.18 | 0.19 | 0.16 |
| Sr1 (-) | 0.12 | 0.10 | 0.14 | 0.16 |
| Sr2 (-) | 0.82 | 0.81 | 0.83 | 0.84 |
| dp1 (-) | 0.11 | 0.16 | 0.17 | 0.14 |
| ypp (-) | 0.48 | 0.55 | 0.58 | 0.58 |

Table 3
Surface topography parameters of Fellows chiselled gear tooth flanks

| Parameters | Gear drive side | Gear drive no side | | |
|---|---|---|---|---|
| | Whole side | Whole side | | Whole side |
| SPa (µm) | 2.56 | 1.54 | 1.15 | 1.25 |
| SPq (µm) | 3.34 | 1.39 | 1.44 | 1.63 |
| SPt (µm) | 25.4 | 14.50 | 8.58 | 10.45 |
| SPsk (-) | 1.08 | -0.25 | -0.75 | -0.99 |
| SPku (-) | 4.37 | 3.52 | 3.36 | 4.23 |
| SPmmr ($mm^3/mm^{2)}$ | 0.0082 | 0.0080 | 0.0050 | 0.0072 |
| SPmvr ($mm^3/mm^{2)}$ | 0.0172 | 0.0056 | 0.0033 | 0.0035 |
| SPds ($pks/mm^{2)}$ | 1888 | 1393 | 1435 | 1360 |
| SPtr (-) | 0.86 | 0.28 | 0.34 | 0.45 |
| SPal (µm) | 509 | 369 | 263 | 330 |
| SP isotropy (%) | 86 | 28 | 34 | 45 |
| SPfd (-) | 2.15 | 2.16 | 2.16 | 2.13 |
| SPΔq (µm /µm) | 0.0415 | 0.0381 | 0.0370 | 0.0391 |
| SPsc (1/µm) | 0.0090 | 0.0086 | 0.0084 | 0.0088 |
| SPdr (%) | 0.086 | 0.072 | 0.070 | 0.074 |
| SPbi (-) | 0.33 | 0.90 | 1.29 | 1.52 |
| SPci (-) | 2.06 | 1.57 | 1.14 | 1.16 |
| SPvi (-) | 0.07 | 0.14 | 0.15 | 0.18 |
| SP(1/fx) (µm) | 133 | 182 | 222 | 250 |
| SP(1/fy) (µm) | 111 | 62 | 77 | 87 |
| SPΔax (µm /µm) | 0.0075 | 0.0049 | 0.0049 | 0.0047 |
| SPΔay (µm /µm) | 0.0157 | 0.0176 | 0.0159 | 0.0196 |
| SPqsum (µm) | 0.29 | 0.43 | 0.41 | 0.43 |
| SPscx (1/µm) | 0.0020 | 0.0013 | 0.0012 | 0.0012 |
| SPscy (1/µm) | 0.0036 | 0.0032 | 0.0032 | 0.0036 |
| SPpq/SPt (-) | 0.23 | 0.11 | 0.13 | 0.13 |
| SPvq/SPt (-) | 0.12 | 0.21 | 0.17 | 0.26 |
| SPmq (-) | 0.06 | 0.75 | 0.56 | 0.73 |
| SPpk/SPt (-) | 0.22 | 0.17 | 0.19 | 0.23 |
| SPvk/SPt (-) | 0.29 | 0.24 | 0.23 | 0.27 |
| SPk/SPt (-) | 0.10 | 0.18 | 0.32 | 0.20 |
| Sr1 (-) | 0.11 | 0.08 | 0.08 | 0.12 |
| Sr2 (-) | 0.80 | 0.81 | 0.87 | 0.82 |
| dp1 (-) | 0.08 | 0.16 | 0.31 | 0.17 |
| ypp (-) | 0.50 | 0.63 | 0.59 | 0.58 |

Table 4
Expected values of primitive profile of gear drive side and gear drive no side obtained with a Form Talysurf profilometer

| Parameters | Hobbing | | | | Fellows chiselling | | | |
|---|---|---|---|---|---|---|---|---|
| | Profile | | Helix | | Profile | | Helix | |
| | Gear drive side | Gear drive no side | Gear drive side | Gear drive no side | Gear drive side | Gear drive no side | Gear drive side | Gear drive no side |
| Pa (μm) | 1.60 | 1.23 | 2.12 | 0.93 | 0.44 | 0.38 | 0.41 | 0.24 |
| Pq (μm) | 1.94 | 1.57 | 2.78 | 1.08 | 0.55 | 0.46 | 0.48 | 0.32 |
| Pt (μm) | 9.74 | 9.22 | 13.23 | 4.96 | 3.50 | 2.91 | 2.32 | 1.68 |
| PzJIS (μm) | 6.89 | 6.05 | 8.28 | 3.39 | 2.42 | 2.21 | 1.12 | 0.69 |
| Pp (μm) | 5.17 | 5.16 | 5.58 | 2.22 | 1.72 | 1.51 | 0.93 | 0.85 |
| Pp/Pt (-) | 0.53 | 0.56 | 0.42 | 0.45 | 0.49 | 0.52 | 0.40 | 0.51 |
| Psk (-) | 0.57 | 0.21 | -0.44 | -0.16 | 0.07 | 0.14 | -0.36 | -0.04 |
| Pku (-) | 2.68 | 3.18 | 2.93 | 2.08 | 2.94 | 2.62 | 2.28 | 3.08 |
| Pk (μm) | 5.16 | 2.63 | 6.62 | 2.93 | 1.33 | 0.99 | 1.14 | 0.54 |
| Ppk (μm) | 2.34 | 3.04 | 1.98 | 0.69 | 0.70 | 0.79 | 0.30 | 0.56 |
| Pvk (μm) | 2.03 | 1.75 | 3.57 | 1.24 | 0.74 | 0.67 | 0.65 | 0.27 |
| Mr1 (%) | 12 | 14 | 12 | 10 | 10 | 12 | 11 | 16 |
| Mr2 (%) | 88 | 84 | 86 | 86 | 90 | 85 | 85 | 81 |
| Ppq (μm) | 0.88 | 1.84 | 0.79 | 0.60 | 0.52 | 0.47 | 0.28 | 0.24 |
| Pvq (μm) | 1.75 | 1.48 | 2.91 | 0.99 | 0.56 | 0.42 | 0.46 | 0.32 |
| Pmq (%) | 2 | 17 | 5 | 7 | 3 | 55 | 15 | 2 |
| PS (μm) | 14 | 13 | 20 | 10 | 10 | 7 | 6 | 4 |
| PSm (μm) | 176 | 86 | 222 | 182 | 46 | 48 | 111 | 89 |
| PΔq (μm/μm) | 0.0456 | 0.0660 | 0.0398 | 0.0314 | 0.05663 | 0.0534 | 0.0349 | 0.0182 |
| PΔa (μm/μm) | 0.0286 | 0.0318 | 0.0216 | 0.0185 | 0.0216 | 0.0328 | 0.0127 | 0.0099 |
| Pλq (μm) | 268 | 150 | 439 | 216 | 1.66 | 54 | 86 | 111 |
| P(1/f) (μm) | 105 | 70 | 135 | 132 | 61 | 23 | 50 | 43 |
| P(β0.1) (μm) | 301 | 218 | 819 | 608 | 31 | 165 | 1016 | 982 |
| $Rq^2$ ($\mu m^2$) | 3.88 | 2.46 | 7.73 | 1.17 | 134 | 0.22 | 0.23 | 0.10 |
| Psc3 (1/μm) | 0.1687 | 0.1083 | 0.0880 | 0.0762 | 0.30 | 0.0932 | 0.0595 | 0.0519 |
| Pds3 (1/mm) | 96.6 | 175.4 | 135.0 | 147.3 | 0.0991 | 224.4 | 150.8 | 159.8 |
| $P\delta^*$ (μm) | 0.11 | 0.08 | 0.09 | 0.02 | 0.09 | 0.07 | 0.03 | 0.02 |
| Kα (-) | 0.50 | 0.52 | - | - | 0.23 | 0.54 | - | - |
| $G^2(\lambda)$ ($\mu m^2$) | 3.7745 | 2.8821 | 8.4065 | 1.4093 | 0.2856 | 0.2349 | 0.1949 | 0.1208 |

Table 5
Expected parameter values of profile and line of technological surface (separated from measurement with a Talyscan 150 profilometer) and modelling flank surface of teeth getting out of hobbing

| Profile parameters | Technological surface | | | | Modelling surface | | | |
|---|---|---|---|---|---|---|---|---|
| | Helix | | Profile | | Helix | | Profile | |
| | Primiti-ve profile | Roug-hness profile | Primiti-ve profile | Roug-hness profile | Primiti-ve profile | Roug-hness profile | Primiti-ve profile | Roug-hness profile |
| Pa, Ra (μm) | 3.17 | 0.65 | 2.45 | 0.93 | 2.00 | 1.95 | 0.25 | 0.02 |
| Pq, Rq (μm) | 3.84 | 1.04 | 3.34 | 1.26 | 2.33 | 2.27 | 0.29 | 0.03 |
| Pt, Rt (μm) | 14.70 | 7.89 | 18.21 | 7.38 | 7.81 | 7.66 | 1.00 | 0.16 |
| Psk, Rsk (-) | -0.63 | -0.52 | -0.39 | -0.41 | 0.64 | 0.66 | -0.10 | 0.11 |
| Pku, Rku (-) | 2.41 | 8.28 | 4.01 | 4.36 | 2.17 | 2.20 | 1.78 | 3.12 |
| PSm, RSm (μm) | 850 | 232 | 425 | 212 | 2000 | 2000 | 1346 | 154 |
| 1/fx, 1/fy (μm) | 366 | 135 | 284 | 135 | 337 | 613 | 369 | 158 |
| Pal, Ral (μm) | 99 | 84 | 84 | 82 | 897 | 144 | 2300 | 37 |
| PΔa, RΔa (μm/μm) | 0.0185 | 0.0159 | 0.0321 | 0.0278 | 0.0072 | 0.0070 | 0.0017 | 0.0014 |
| PΔq, RΔq(μm/μm) | 0.0339 | 0.0304 | 0.0426 | 0.0381 | 0.0084 | 0.0082 | 0.0019 | 0.0016 |
| Psc, Rsc (1/μm) | 0.0010 | 0.0007 | 0.0020 | 0.0018 | 0.0007 | 0.0007 | 0.0001 | 0.0001 |
| Psd, Rsd (1/mm) | 70.6 | 117.6 | 117.6.1 | 152.9 | 5.5 | 5.5 | 158.0 | 24.7 |
| Pwp, Rwp (μm) | 0.39 | 0.28 | 0.20 | 0.47 | 0.25 | 0.23 | 0.45 | 0.12 |
| Ppq/Pt, Rpq/Rt (-) | 0.11 | 0.18 | 0.17 | 0.16 | 0.36 | 0.37 | 0.03 | 0.26 |
| Pvq/Pt, Rvq/Rt (-) | 0.32 | 0.28 | 0.32 | 0.31 | 0.20 | 0.19 | 0.30 | 0.22 |
| Pmq, Rmq (-) | 0.42 | 0.98 | 0.64 | 0.72 | 0.57 | 0.56 | 0.24 | 0.10 |
| Ppk/Pt, Rpk/Rt (-) | 0.03 | 0.25 | 0.23 | 0.23 | 0.58 | 0.18 | 0.45 | 0.26 |
| Pvk/Pt, Rvk/Rt (-) | 0.46 | 0.25 | 0.23 | 0.22 | 0.95 | 0.95 | 0.14 | 0.18 |
| Pk/Pt, Rk/Rt (-) | 0.56 | 0.15 | 0.31 | 0.34 | 0.48 | 0.48 | 0.63 | 0.49 |
| PMr1, RMr1 (-) | 0.05 | 0.13 | 0.16 | 0.16 | 0.21 | 0.06 | 0.19 | 0.12 |
| PMr2, RMr2 (-) | 0.81 | 0.82 | 0.87 | 0.87 | 0.96 | 0.96 | 0.90 | 0.95 |
| Pdp1, Rpd1 (-) | 0.53 | 0.13 | 0.28 | 0.30 | 0.46 | 0.46 | 0.56 | 0.46 |
| Pypp, Rypp (-) | 0.28 | 0.46 | 0.41 | 0.59 | 0.79 | 0.79 | 0.60 | 0.55 |
| Kα, - | - | - | 0.24 | 0.32 | - | - | 0.98 | ~1.00 |

Table 6
Expected parameter values of profile and line of technological surface (separated from measurement with a Talyscan 150 profilometer) and modelling flank surface of teeth getting out of Fellows chiselling

| Profile parameters | Technological surface | | | | Modelling surface | |
|---|---|---|---|---|---|---|
| | Helix | | Profile | | Profile | |
| | Primitive profile | Roughness profile | Primitive profile | Roughness profile | Primitive profile | Roughness profile |
| Pa, Ra (μm) | 0.34 | 0.14 | 1.12 | 0.46 | 0.46 | 0.04 |
| Pq, Rq (μm) | 0.39 | 0.20 | 1.34 | 0.56 | 0.80 | 0.05 |
| Pt, Rt (μm) | 1.70 | 1.48 | 5.29 | 2.83 | 3.51 | 0.32 |
| Psk, Rsk (-) | -0.12 | -0.86 | -0.48 | -0.02 | -0.31 | -0.40 |
| Pku, Rku (-) | 1.94 | 6.34 | 2.29 | 2.84 | 2.33 | 4.69 |
| PSm, RSm (μm) | 1280 | 160 | 284 | 213 | 1346 | 128 |
| 1/fx, 1/fy (μm) | 182 | 80 | 162 | 107 | 392 | 374 |
| Pal, Ral (μm) | - | 142 | - | 250 | - | 34 |
| PΔa, RΔa (μm/μm) | 0.0084 | 0.0078 | 0.0047 | 0.0070 | 0.0073 | 0.0063 |
| PΔq, RΔq μm/μm) | 0.0096 | 0.0091 | 0.0058 | 0.0084 | 0.0105 | 0.0073 |
| Psc, Rsc (1/μm) | 0.0004 | 0.0003 | 0.0007 | 0.0007 | 0.0001 | 0.0001 |
| Psd, Rsd (1/mm) | 305.9 | 388.2 | 58.8 | 58.8 | 50.0 | 162.5 |
| Pwp, Rwp (μm) | 0.51 | 0.41 | 0.29 | 0.30 | 0.19 | 0.33 |
| Ppq/Pt, Rpq/Rt (-) | 0.13 | 0.11 | 0.04 | 0.07 | 0.12 | 0.25 |
| Pvq/Pt, Rvq/Rt (-) | 0.20 | 0.50 | 0.27 | 0.21 | 0.23 | 0.17 |
| Pmq, Rmq (-) | 0.21 | 0.78 | 0.29 | 0.20 | 0.34 | 0.08 |
| Ppk/Pt, Rpk/Rt (-) | 0.17 | 0.17 | 0.10 | 0.32 | 0.09 | 0.28 |
| Pvk/Pt, Rvk/Rt (-) | 0.30 | 0.28 | 0.05 | 0.17 | 0.36 | 0.25 |
| Pk/Pt, Rk/Rt (-) | 0.58 | 0.20 | 0.69 | 0.44 | 0.52 | 0.21 |
| PMr1, RMr1 (-) | 0.09 | 0.14 | 0.03 | 0.19 | 0.08 | 0.15 |
| PMr2, RMr2 (-) | 0.86 | 0.83 | 0.85 | 0.92 | 0.85 | 0.84 |
| Pdp1, Rpd1 (-) | 0.54 | 0.18 | 0.65 | 0.39 | 0.49 | 0.18 |
| Pypp, Rypp (-) | 0.40 | 0.33 | 0.28 | 0.50 | 0.34 | 0.47 |
| Kα, - | - | - | 0.92 | 0.87 | 1 | 1 |

## 7. Result analysis

Hobbing, as roughing of the investigated cylindrical wheels, results in far worse wheel accuracy than the finishing chiselling by Fellows method. However, it does not concern the total cumulative deviation of wheel pitch $F_p$. It has a smaller value for hobbing (Table 1). It follows from the final formation of the tooth flanks with one hob blade. The precision of tooth adherence and side play, characterized with $F_r$ and $E_s$ deviations, respectively, is greater for Fellows chiselling. The two methods of gear tooth generation ensure better precision of the tooth flanks getting out of machining than those getting into machining (Table 1). It was found after having examined all the tangential composite deviations $F_a$, $F_\beta$, $f_{pt}$ and $F_p$.

Flaws in the tooth flanks after hobbing along the tooth face width follow from the kinematic ratio of the technological gear. Numerous scratches of the tooth flanks getting into hobbing or chiselling can be accounted for as due to unfavourable conditions of chip formation and flow.

A great roughness height is characteristic for tooth flanks getting into machining than those getting out (Tables 2 and 3). A similar situation is with roughness spacing, which is evidenced by the correlation function length. Slope, peak curvature and their density seem to confirm this as well. Another characteristic feature of the surfaces getting into machining is greater summit heights of small material ratio and smaller valley depths. The isotropy of the tooth flanks getting into machining is particularly great for Fellows chiselling. It follows from the texture aspect ratio of the surface SPtr and the isotropy index SPizo. It has also been proved by the results of the research on the roughness height, spacing, curvature and cumulated power spectral density of surface profiles for tooth contour and helix profile (Table 4). However, it has to be pointed out that a chiselled tooth flank is more anisotropic than hobbed one. The material ratio curve calculated from the surface profile has a far bigger kernel depth Pk and bigger standard valley deviation Pvq for tooth flanks getting into machining compared to those getting out. The above presented tooth flank features make the flanks of used cylindrical gear wheels, especially those chiselled off by Fellows method, be machined as flanks getting out of machining.

Hobbed and Fellows chiselled gear wheels are characterized by a greater surface roughness height of the tooth root than the tooth point. It is evidenced by the following calculated parameter values of arithmetic and quadratic means, SPa and SPq, of surface deviation as well as the total surface height SPt (Tables 3, 4). Quadratic means of the roughness summit heights SPqsum also have greater values on the tooth root than on its point. The tooth root surface compared to its point has a higher value of mean material volume ratio SPmmr and mean void volume ratio SPmvr. The conclusions are confirmed by the index values of the retention of fluid in the surface core SPci and those in the surface valleys SPvi. The height variance of the tooth point and root surface roughness is far greater for hobbing.

Spaces between the surface irregularities in the tooth point area are smaller compared to the root area. It refers to both hobbing and chiselling. It is evidenced by the correlation length SPal. Similar conclusions follow from the characteristic wave length value of the helix profile SP(1/fx), and tooth profile SP(1/fy), Figs 5 and 6, Tables 2 and 3). Power spectral density is greater for the root surface of both hobbed and Fellows chiselled teeth than for the tooth point. The presented surface toughness height and spacing change tendencies make the slopes in these areas correspondingly similar. Roughness summits density is greater on the tooth point surface than on its root area. The fractal dimension SPdf is always greater for the tooth point area. The surface ordinate distribution defined by the surface skewness ratio SPsk and surface kurtosis ratio Psku has varied values for the tooth flank areas under consideration. Due to this, the interpretation of PSsk and PSku values is difficult. That results from their sensitivity to accidental summits and valleys of the surface.

The characteristics of the material ratio curve are listed. Its description, based on the secant (parameters SPpk, SPvk, SPk, Mr1 and Mr2), is varied for tooth point and root areas in the case of both hobbing and chiselling. The description of the material ratio curve in the probability density system is univocal. Standard deviations of roughness summits of tooth point and root surfaces SPpq are correspondingly the same. However, standard valley deviations SPvq are always greater for the root area. The material ratio SPmg has also bigger values for the tooth root surface than for the tooth point. Considering the material ratio curve approximated with three parameters it must be stated that its slope at the inflexion point dp1 is greater for the point area than for the root area. In turn, the ordinate of the point is appropriately similar for the areas analysed. The ordinate is close to the mode of the surface ordinate distribution density.

The tooth flank surface is oriented, particularly after chiselling. The surface texture is well characterized by the contour map and the angular plot of the power spectral density function (Figs 5 and 6). The flanks of a hobbed gear wheel have a more oriented root, whereas the flanks of a chiselled gear have a more oriented tooth point. That is demonstrated by the values of the texture aspect ratio of the surface SPtr and those of the isotropy index SPizo. I also follows from the cumulated and arbitrary power spectral density along the helix and tooth profile.

The presentation of the characteristics of the topography of the tooth flanks on the basis of the profile and helix is given in Tables 5 and 6. The given values refer to both a technologically worked surface coming out of the machining and a model surface for hobbing and Fellows chiselling. The heights of the surface irregularities along the pitch helix of a hobbed gear wheel are bigger than in the tooth profile. However the flanks of the Fellows method chiselled teeth have a smaller irregularity height along the helix than in the profile. It is confirmed by the $K_\alpha$ coefficient value. The plots of the primitive profile autocorrelation function of the tooth flanks, on the profile and along the helix were qualified as characterizing the periodical profile of a great wave length (Figs 7 and 8). It is confirmed by the power spectral density characterized with a long range of wave length. The strongest are the waves of primitive profiles of 650 μm and those of roughness profiles of about 350 μm. The values of these characteristic wave lengths of profiles for model tooth flank surfaces are about 1200 μm and 150 μm, respectively. The application of Gauss's filter made it possible to get rid of the primitive surface and profile error.

Despite a small module of wheel rims, differences in respect of height and spacing of the roughness profile of tooth contour were found between the area below the peak cylinder and the area above the active cylinder. Slight differences were found while comparing the height of the helix profile of real wheels with that of modelling wheels. The flank of hobbed gear teeth has roughness heights of a few micrometers. The parameters of the total height of the helix profile are Pt=14.70 μm, Rt=7.89 μm. In the surface model, the irregularities are Pt=7.81 μm, Rt=7.66 μm (Fig. 7, Table 5). The value calculated from equation (1) is $\delta_x$=5.53 μm. The pitch helix of a chiselled real wheel shows irregularities of total height Pt=1.70 μm, Rt=1.48 μm. In the model of the tooth flank the pitch helix is a straight line (Fig. 8, Table 6). Slightly greater differences follow from the comparison real and model wheels along the profile height. For hobbing, the tooth profile for a real wheel surface is Pt=3.34μm, Rt=1.26 μm. The values are Pt=0.29 μm, Rt=0.03 μm for the wheel model. The value calculated from equation (2) is $\delta_y$=0.09 μm. The flank surface of a chiselled gear wheel has a tooth profile of heights Pt=5.29 μm, Rt=2.83 μm. The corresponding values for the model surface are Pt=3.51 μm, Rt=0.32 μm. Roughness profiles of the tooth contour of the model surface have small spacing changes, though. It was found that primitive profiles and flank roughness profiles are periodical or quasiperiodic. It refers to both tooth contour and pitch helix profile of hobbed and Fellows chiselled wheels as well as to wheels obtained through simulation. A comparison of the values of the parameters of the analysed real and model profiles can be found in Tables 5 and 6. Differences in respect of the height of profile irregularities, their spacing and slope as well as the peak heights, their density and curvature were found. It confirms a great effect of the physical phenomena of machining on the tooth flank surface topography parameters. It also follows from the mistakes made while analyzing the length difference of the probing step, quantization level and the assumed associated integral elements.

Making use of a conregular model of surface topography, similar profile lengths of rising and falling irregularities profiles were found. It refers to tooth profiles of both real and model wheels (Figs 9 and 10). The profile of the helix of machined wheels is asymmetric. Simultaneous slope of the flank profiles for rising irregularities is different from falling ones. The technological, worked flanks show considerable differences between these slopes on the helix. However, model surfaces have the same slope of rising and falling surface irregularities of the helix. Contour profiles along the tooth height of technologically worked and a model surface have a similar slope diversification. The occurring slope differences of helix profile for rising and falling segments were thought to be due to the effect of physical phenomena of the machining process.

## 8. Conclusions

The presented methodics of modelling cylindrical gear flanks allows a precise definition of the stereometric shape and surface topography. The methodics is particularly useful as it enables obtaining tooth profiles of any degree of complication. Models obtained in this way may be analysed along the tooth contour as well as along the helix at any section. Such an analysis is difficult, and often impossible when the surface is characterized with mathematical equations.

Assuming that the turning step corresponds to the number of the tool cutting edge contacts necessary to make one tooth flank during the real machining of a toothed wheel, it is possible, already at the CAM elaboration stage, to define the precision of gear generation resulting from the machining kinematics. Supposing that the machining simulation runs under ideal conditions, it is possible to determine the effect of tool setting and the machine-tool-chuck-object-tool system on gear tooth making precision. Model surface of tooth flanks allows analyzing monominal flank deviations, as well as radial composite and runout ones.

The experimental results confirmed the correctness of the worked out model of the tooth flanks of a cylindrical gear wheel as to the character of the surface topography after hobbing and Fellows chiselling. It was found that the roughness height and roughness spacing were smaller for the tooth point surface than for the root surface of hobbed and Fellows chiselled teeth. The flank tooth surface of a gear wheel made by Fellows chiselling method is anisotropic and strongly oriented along the helix. The orientation of the flanks after hobbing is far smaller. Stereoscopic pictures show the occurrence of numerous flaws in both hobbed and chiselled tooth flanks.

Hobbing gives far worse accuracy of a cylindrical gear wheel than chiselling by Fellows method. Hobbing, however, makes it possible to achieve a smaller total cumulative pitch deviation $F_p$. It results from the tooth flank being finally precision shaped with one hob blade. The precision of the gear tooth flanks getting out of the machining process is higher than those getting into it. It was found on the basis of all the examined tangential composite deviations $F_a$, $F_\beta$, $f_{pt}$ and $F_p$.

It also follows from the analysis of the measurements obtained with two and three dimensional profilometers that the primitive and roughness profiles along at the tooth height and helix along tooth width are quasiperiodic. The above refers to all the tooth flanks - hobbed, chiselled and model ones obtained by the generation simulation method.

Taking into account the conregular model of the topography of the generated surface it was found that the profile lengths of the roughness rise and fall are similar and that the slope of the tooth profile and helix vary. Model tooth flanks related to hobbing and Fellows chiselling have symmetrical and identical slope of the flanks of tooth roughness.

Tooth flanks getting into hobbing and chiselling have bigger roughness height, spacing, slope, peak curvature and their density than those getting out of machining. They also show greater summit height of low material ratio and smaller valley depth of the surface. It refers particularly to gear teeth chiselled by Fellows method. In the case of model gears there is no difference between flanks coming into and getting out of machining. Model toothed wheels also show smaller tooth profile and pitch helix deviation values than real gear wheels. It is due to the fact that during the simulation it is impossible to allow for extra physical phenomena that accompany real machining.